\documentclass[%
 reprint,
 amsmath,amssymb,
aps,
nofootinbib,
nobibnotes,
]{revtex4-1}

\usepackage{graphicx}
\usepackage{dcolumn}
\usepackage{bm}
\usepackage{booktabs}
\usepackage{array}
\usepackage{placeins}
\usepackage[caption=false]{subfig}

\newcommand\blfootnote[1]{%
  \begingroup
  \renewcommand\thefootnote{}\footnote{#1}%
  \addtocounter{footnote}{-1}%
  \endgroup
}

\begin{document}


\title{Valence instability across magnetostructural transition in USb$_2$ }

\author{Z. E. Brubaker$^{1,2,3}$, Y. Xiao$^{4}$, P. Chow$^{4}$, C. Kenney-Benson$^{4}$, J. S. Smith$^{4}$, H. Cynn$^{1}$, C. Reynolds$^{1}$, N. P. Butch$^{5,6}$, R. J. Zieve$^{2}$ and J. R. Jeffries$^{1}$\\
$^{1}$ Lawrence Livermore National Laboratory, Livermore, California 94550\\
$^{2}$ Physics Department, University of California, Davis, California 95616\\
$^{3}$ Oak Ridge National Laboratory, Oak Ridge, Tennessee 37831\\
$^{4}$ HP-CAT, X-ray Science Division, Argonne National Laboratory, Argonne, Illinois 60439\\
$^{5}$ NIST Center for Neutron Research, National Institute of Standards and Technology, Gaithersburg, Maryland 20899\\
$^{6}$ Center for Nanophysics and Advanced Materials, Department of Physics, University of Maryland, College Park, Maryland 20742\\
}
\date{\today}

\begin{abstract}


We have performed pressure dependent X-ray diffraction and resonant X-ray emission spectroscopy experiments on USb$_2$ to further characterize the AFM-FM transition occuring near 8~GPa. We have found the magnetic transition coincides with a tetragonal to orthorhombic transition resulting in a 17\% volume collapse as well as a transient \textit{f}-occupation enhancement. Compared to UAs$_2$ and UAsS, USb$_2$ shows a reduced bulk modulus and transition pressure and an increased volume collapse at the structural transition. Except for an enhancement across the transition region, the \textit{f}-occupancy decreases steadily from 1.96 to 1.75.
\end{abstract}
\pacs{Valid PACS appear here}
\maketitle


\section{Introduction}
\blfootnote{Notice:  This manuscript has been authored by UT-Battelle, LLC, under contract DE-AC05-00OR22725 with the US Department of Energy (DOE). The US government retains and the publisher, by accepting the article for publication, acknowledges that the US government retains a nonexclusive, paid-up, irrevocable, worldwide license to publish or reproduce the published form of this manuscript, or allow others to do so, for US government purposes. DOE will provide public access to these results of federally sponsored research in accordance with the DOE Public Access Plan (http://energy.gov/downloads/doe-public-access-plan).}

\textit{F}-electron quantum materials exhibit a variety of electronic and magnetic phenomena---such as heavy fermi liquids, mixed valence states, long range magnetic ordering and superconductivity---that are intrinsically coupled to \textit{f}-electron hybridization. \cite{Coleman,Stewart} Chemical substitution, pressure, and magnetic field can tune the \textit{f}-electron hybridization, thus allowing new magnetic states to emerge. In U-based compounds, magnetic order is particularly sensitive to the U-U spacing, generally  not forming magnetically ordered states with U-U spacing below the Hill limit of 3.5 \AA .\cite{Hill} Consequently, a variety of U-compounds exhibit exotic phase diagrams as the U--U spacing is tuned, making them promising candidates to understand the relationship of structure and magnetism.\cite{UCompounds} Among the U-compounds, the uranium dipnictides (UX$_2$) exhibit some of the highest magnetic transition temperatures, accompanied by particularly large U-U separations, creating a promising environment for extensive pressure-dependent studies.

\begin{figure}[!htbp]
	\centering
	\includegraphics[width=\linewidth]{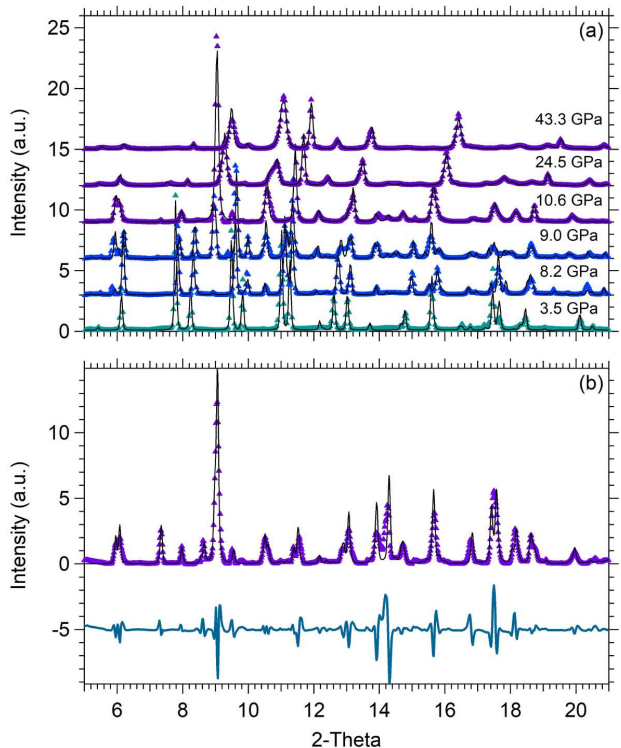}
	\caption{(a) Representative background-subtracted PXRD patterns (markers) and fit (solid lines). The P=8.2 GPa and 9 GPa data show the coexistence of each phase, and the bump just below 10 degrees and at 15 degrees in the 10.6 GPa data correspond to the last remaining peaks of the low pressure phase. (b) Distinct PXRD measurement at P=10.5 GPa used to determine the high-pressure phase. The solid black line shows the resulting LeBail fit and accounts for each peak in the pattern, and the blue ticks correspond to the reflections for the Pmm2 structure; the small bump below 10 degrees is due to the low pressure phase, and the bump near 15 degrees is an artifact of the measurement that showed up at all pressures for this measurement. The solid blue line shows the difference curve and is offset by $-5$.}
	\label{fig:PXRD}
\end{figure}
 
The UX$_2$ (X=As, Sb, Bi) compounds crystallize in the tetragonal anti-Cu$_2$Sb type structure (P4/nmm). Although UP$_2$ was originally thought to also crystallize in this structure, more recent structural studies have shown that the diffraction pattern is better described using the I4mm space group with three unique U-sites. \cite{UP2_Structure} The antiferromagnetic ordering temperature (T$_N$) decreases with increasing atomic size (and thus U-U spacing) in contrast to the Hill scenario for U-compounds, and shows ordering temperatures of T$_N$=273 K, 203 K, and 183 K for UAs$_2$, USb$_2$, and UBi$_2$, respectively. \cite{Henkie_USb2, UBi2, Hill} Owing to its distinct crystal structure, UP$_2$ does not follow this trend, and shows an ordering temperatures of 203 K, though doping studies in the U(P,As)$_2$ system have shown a steady increase in T$_N$ as the As content is increased. \cite{Henkie} 

\begin{table*}[!htbp]
\centering
\renewcommand\arraystretch{1.2}
\caption{Lattice parameters (\AA) at ambient-pressure (LP) and 45 GPa (HP) and comparison of bulk modulus and its derivative (B and B$^\prime$), transition pressure (P$_c$), and volume contraction for UP$_2$, UAsS, UAs$_2$, and USb$_2$. \cite{HTHP} Although the bulk modulus and $\Delta$V of UP$_2$ are listed in the literature, they were calculated assuming an ambient pressure P4/nmm structure and are thus unreliable. \cite{HTHP}}
\label{table:UX2}
\begin{ruledtabular}
\begin{tabular}[t]{lccccccccc}
Compound&a$_{LP}$&c$_{LP}$&a$_{HP}$&b$_{HP}$&c$_{HP}$&B$_{LP}$ (GPa)&B$^\prime_{LP}$&P$_c$ (GPa)&$\Delta V$ (\%)\\

\hline

UP$_2$&5.39&15.56&unknown&unknown&unknown&unknown&unknown&22&unknown\\
UAsS&3.88&8.16&unknown&unknown&unknown&105&3.7&46&7\\
UAs$_2$&3.96&8.12&3.56&6.4&8.53&101&4.7&15&7\\
USb$_2$&4.29&8.76&4.13&6.61&7.46&69&4.7&8&17\\

\end{tabular}
\end{ruledtabular}
\end{table*}

Pressure-dependent transport measurements performed on USb$_2$ indicate that T$_N$ approaches that of UAs$_2$, but an abrupt AFM--FM transition occurs near P=8 GPa, reducing the ordering temperature by over 100 K. \cite{USb2} The AFM-FM transition has been the subject of several recent theoretical calculations, though from an experimental standpoint, only little is known about this transition. \cite{USb2_2018,USb2_2019} Based on structural tetragonal to orthorhombic transitions in the related UAsS, UP$_2$, and UAs$_2$ structures at 46 GPa, 22 GPa, and 15 GPa, respectively, it stands to reason that USb$_2$ may undergo a coupled magnetostructural transition. \cite{HTHP} The U \textit{f}-occupancy, n$_f$, may also be affected by such a magnetostructural transition and could play a fundamental role in dictating the magnetic structure. Thus, to provide a more complete description of the pressure-induced magnetic transition observed in USb$_2$, we have performed pressure dependent powder X-ray diffraction (PXRD) and resonant X-ray emission spectroscopy (RXES) experiments at room temperature. The structure undergoes a tetragonal to orthorhombic transition that conincides with the magnetic transition. RXES measurements suggest a gradual suppression of \textit{f}-occupancy with pressure, though an unusual temporary enhancement of n$_f$ occurs in the transition region.
 
\begin{figure}[hbtp]
	\centering
	\includegraphics[width=\linewidth]{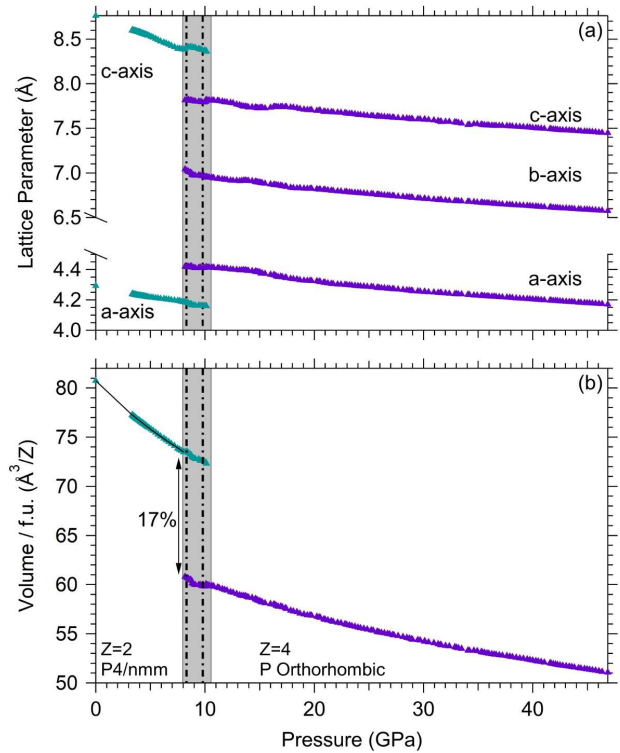}
	\caption{Calculated lattice parameters and volume for USb$_2$. The gray rectangle corresponds to the transition width found from the structural data, and the dashed black line is the transition width measured from transport measurements. \cite{USb2} The solid line in (b) represents a fit to the Birch--Murnaghan equation of state up to 8 GPa, and yields values of B=68.7 GPa and B$^\prime$=4.7. Error bars are smaller than the markers and have been omitted for clarity.}
	\label{fig:structural_results}
\end{figure}

\section{Experimental Methods}

USb$_2$ was grown by self flux as described previously. \cite{USb2} Powder X-ray diffraction (PXRD) measurements were performed at sector 16-IDB of the Advanced Photon Source (APS) using a 30 keV X-ray beam. Powdered USb$_2$ was loaded into a diamond anvil cell (DAC) using a rhenium gasket and pressurized with neon. The DAC was double-encapsulated, and used kapton and mylar for the second layer of encapsulation. The pressure was determined from copper powder mixed in with the sample and controlled with a gas-driven membrane. All patterns were analyzed using fit2D and GSAS-II. \cite{Fit2D,GSASII} The instrument parameters were determined from a CeO$_2$ calibration sample and were held constant for all refinements; only the lattice parameters, intensities, preferred orientation parameters, and peak broadening due to pressure were allowed to vary. The PXRD measurements were performed with two distinct samples at the APS. The peak intensities vary between these patterns because each suffers from different single crystal peaks (see appendix).

Resonant X-ray emission spectroscopy (RXES) at the U L-III absorption edge and L-alpha emission line were performed at sector 16-IDD of the APS.  Small crystals, one to two rubies, and mineral oil were loaded into a DAC using a beryllium gasket. As with the PXRD measurements, the DAC was doubly encapsulated for radiological safety. The pressure was determined via ruby fluorescence and controlled with a gas-driven membrane. \cite{Mao} The pressure was measured before and after each RXES scan, and the averaged pressure is presented herein; the error represents the maximum and minimum measured pressures. 

A self-absorption correction was not applied to the RXES data. This correction generally assumes an ``infinitely'' thick sample, i.e. $\mu(E)$z $>>$ 1, where $\mu$ is the total absorption and z is the sample thickness. \cite{FLUO, Booth_Pu} This approximation is not valid for the scale of samples used in DACs, where the typical sample thickness is approximately 10 $\mu$m. 
	
Throughout this paper, error bars correspond to one standard deviation unless otherwise noted. Identification of commercial equipment does not imply recommendation or endorsement by the National Institute of Standards and Technology.
	
\section{Results}

\subsection{Structure}

Figure \ref{fig:PXRD} shows PXRD patterns at select pressures and the fit for each spectrum. New diffraction peaks begin to emerge near 8 GPa and completely replace the low-pressure peaks just above 10.5 GPa. UAs$_2$ and UAsS, each of which order in the same structure as USb$_2$ at ambient pressure, and UP$_2$, which orders in a similar structure at ambient pressure, undergo a tetragonal to orthorhombic transition under pressure, and the high pressure phase has been assigned the PbCl$_2$ structure, space group Pbnm. \cite{HTHP} This space group, however, fails to account for several of the observed  peaks in the high-pressure phase of USb$_2$. The structural results for UP$_2$, UAs$_2$ and UAsS were determined from energy-dispersive X-ray diffraction over two decades ago, and thus may have (i) lacked the resolution to measure the smaller peaks and (ii) suffered from fluorescence lines obfuscating the diffraction peaks. 

To determine the structure, the diffraction peaks were indexed using GSAS-II. \cite{GSASII} The best fit at P=10.5 GPa---as judged by the residual, peak locations, and cell volume---was obtained with lattice constants of a=4.398 \AA, b=6.944 \AA, and c=7.820 \AA\ using any of the following space groups: P222 (No. 16), P2221 (No. 17), P21221 (No. 18), Pmm2 (No. 25), P21ma (No. 26), and Pmmm (No. 47). Unfortunately, we were unable to identify the atomic positions. This is in part due to the strong single-crystal peaks observed in the 2D patterns, but also because of the low multiplicity of the individual Wyckoff positions and the relatively large number of atoms that need to be placed in this cell with Z=4. In the absence of a theoretical model for the high-pressure structure, we are thus unable to discern the atomic positions and confidently determine the exact space group. That being said, the volume and lattice parameters only depend on the choice of Bravais lattice. As a result, the exact space group of these systems is immaterial to the results discussed herein. We also point out that the Pmm2 space group is a subgroup of the low-pressure P4/nmm phase, and is a promising first structure to investigate. The appendix includes a more complete description of the various attempts to discern the atomic positions. 

Because of these complications, all analysis of the high-pressure phase was performed using the LeBail method. \cite{LeBail} The resulting LeBail fit is shown in Fig. \ref{fig:PXRD}b along with the residual. The residual is somewhat large, but the primary differences arise because of the peakwidths---rather than peak locations---which will not directly influence the inferred volume. We also speculate that UAs$_2$ and UAsS order in the same high-pressure structure as USb$_2$, though the atomic positions would first need to be determined and new PXRD measurements of UAs$_2$ and UAsS would be required to confirm this hypothesis. 

The resulting pressure dependent lattice parameters and volume are shown in Fig. \ref{fig:structural_results}. We have assumed that the number of formula units doubles in the high pressure phase. The low-pressure phase is well described with the Birch--Murnaghan equation of state with values of B=68.7 GPa and B$^\prime$ = 4.7. \cite{BM} At the transition, we observe a surprisingly large volume collapse of $\sim$17\%. These values are compared to those measured for UP$_2$, UAs$_2$, and UAsS in Table \ref{table:UX2}. We note that the structural work performed on UP$_2$ assumed a P4/nmm ambient pressure structure, and thus only the transition pressure is reliable from that analysis. Compared to UAsS and UAs$_2$, USb$_2$ shows a decreasing bulk modulus and transition pressure, and an increasing volume collapse at the structural transition. Following this trend and taking into account that UBi$_2$ orders in the same ambient pressure structure, we suggest that UBi$_2$ could exhibit a structural transition as low as 1--2 GPa.

\subsection{Resonant X-ray Emission Spectroscopy}

\begin{figure}[!tp]
	\centering
	\includegraphics[width=\linewidth]{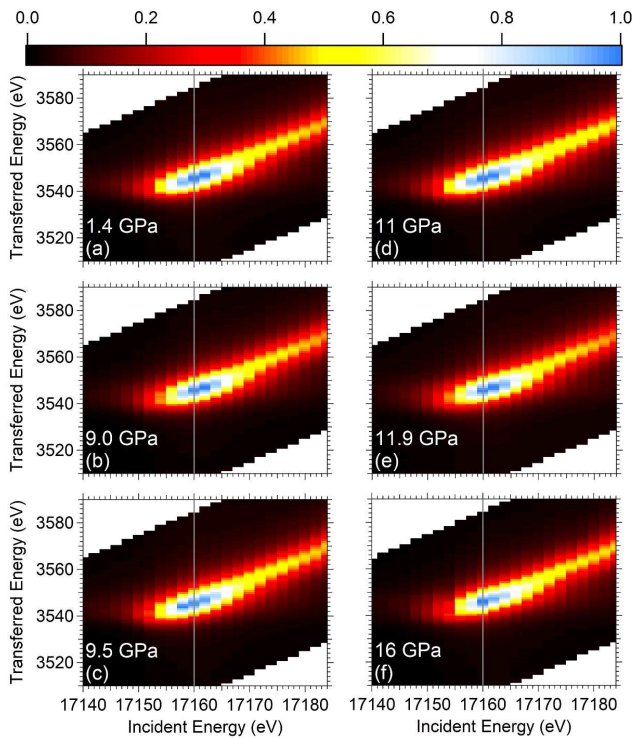}
	\caption{RXES spectra at select pressures. There exists an overall trend of intensity shifting to higher incident/transferred energies with pressure, though from 9.5 GPa to 11 GPa an unusual enhancement of the low-energy intensity occurs. The vertical gray line is a guide to the eye for easier comparisons between pressures. Each spectrum is normalized to a peak intensity of 1.0.}
	\label{fig:RXES_spectra}
\end{figure}

Because of the small energy splitting of the individual U valence states, determining the valence from conventional X-ray absorption spectroscopy (XAS) measurements is difficult. Instead, it is necessary to perform RXES measurements, in which the resolution is set by the 3d$_{5/2}$ orbital ($\sim$4 eV) rather than the 2p$_{3/2}$ orbital ($\sim$7--10 eV). Figure \ref{fig:RXES_spectra} shows the obtained RXES spectra at several of the measured pressures. For U-compounds, the f$^1$, f$^2$, and f$^3$ absorption edges are separated by approximately 4--5 eV, with the f$^3$ absorption edge corresponding to the lowest absorption energy (E$_i$ = 17,156 eV) of these states. As such, in the absence of any obfuscating effects (see discussion), a shift in intensity to higher (lower) energies will correspond to a decrease (increase) in \textit{f}-occupation. 

\begin{figure}[!hbtp]
	\centering
	\includegraphics[width=\linewidth]{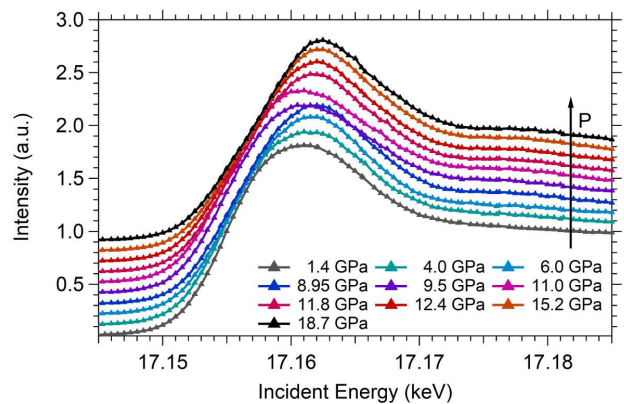}
	\caption{X-ray absorption spectra in partial fluorescence yield mode at select pressures. The peak location shifts to higher incident energies, E$_i$, with increasing pressure, indicating a decrease in \textit{f}-occupation. The abrupt enhancement from 9.5~GPa to 11 GPa is particularly obvious in these scans. Each scan is offset by 0.1 and is normalized to an edge jump of unity. Error bars are smaller than the markers and have been omitted.}
	\label{fig:XAS}
\end{figure}

\begin{table*}[!hbtb]
\centering
\renewcommand\arraystretch{1.2}
\caption{Peak positions (x$_n$), Lorentzian widths ($\Gamma_n$), and valence peak skew factor ($\alpha$) used for RXES analysis. The peak position and width of the lowest energy valence state were allowed to vary a small amount. }
\label{table:Fit_param}
\begin{ruledtabular}
\begin{tabular}[t]{lcccccccc}
x$_3$ (eV)&x$_2$ (eV)&x$_1$ (eV)&x$_F$ (eV)&$\Gamma_3$&$\Gamma_2$&$\Gamma_1$&$\Gamma_F$&$\alpha$\\
\hline
3540$\pm$ 1&3544.2&3548.4&E$_i$ - 13614.5 $\pm$ 0.5&4$\pm$ 0.5&3.1&2.7&5.5&0.3

\end{tabular}
\end{ruledtabular}
\end{table*}

\begin{figure*}[!hbtp]
	\centering
	\includegraphics[width=\linewidth]{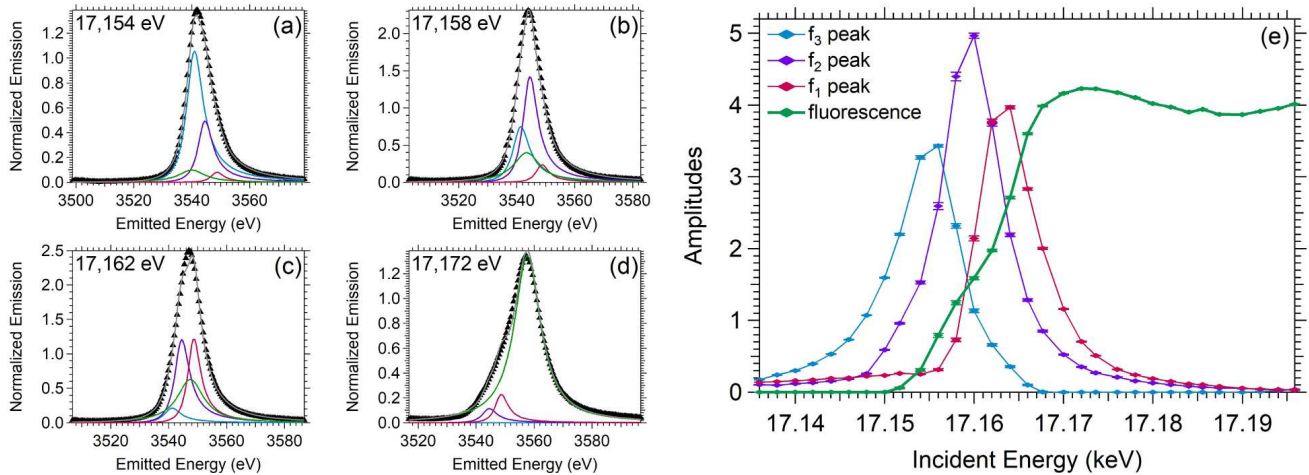}
	\caption{(a--d) Example fits at P=1.4 GPa to the emission scans at incident energies used to determine the line shapes. (e)~Lorentzian amplitudes as a function of pressure for P=1.4 GPa. Error bars from the fitting procedure are included in (e), though they are only discernible from the data points near E$_i$=17.160 keV.}
	\label{fig:amplitudes}
\end{figure*}

Qualitatively, the evolution of n$_f$ can immediately be determined from the measured spectra: n$_f$ decreases with increasing pressure, though it shows an unusual anomaly from 9.5 GPa to 11 GPa, during which n$_f$ is temporarily enhanced. This  transient change is particularly obvious in the X-ray absorption spectra shown in Fig. \ref{fig:XAS}, which are equivalent to constant emitted energy slices of the RXES spectra.

\begin{figure}[hbtp]
	\centering
	\includegraphics[width=\linewidth]{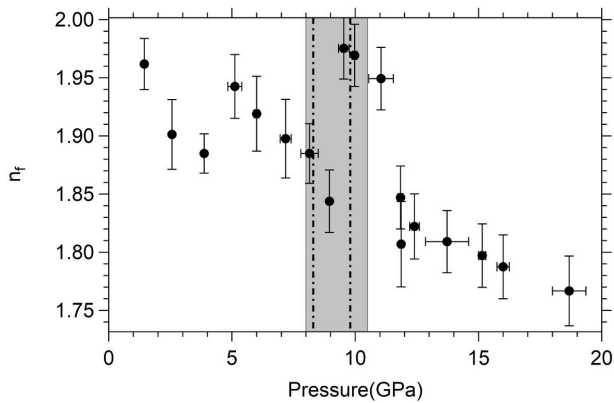}
	\caption{n$_f$ determined via the method proposed by Dallera et al. \cite{Dallera} There is a gradual decrease in n$_f$, though a temporary enhancement occurs across the magnetostructural transition. The gray rectangle corresponds to the transition width found from the structural data and the dashed black lines show the transition width measured from transport measurements. \cite{USb2} An offset of around 1 GPa between the pressure measured in structure and X-ray absorption would be plausible.}
	\label{fig:nf_dallera}
\end{figure}

To provide a more quantitative analysis of n$_f$, we have analyzed each RXES spectra following the procedure proposed by Dallera et al. and used on several U and Pu compounds by Booth et al. \cite{Dallera,Booth} Each emission scan is fit with a skewed Lorentzian (see \cite{Booth}) corresponding to each valence peak (three total), and an additional Lorentzian for the fluorescence peak. The line shapes (i.e., Lorentzian widths) were determined from incident energies just above each of the emission edges, and the skew parameter was held constant for each of the valence peaks for consistency. After determining the line shapes, an average peak position was determined as a function of incident energy near the emission edge. To ensure adequate fits, it was necessary to allow the peak position and width of the f$_3$ peak to vary slightly; previous work allowed the position to vary slightly, but held the width constant. \cite{Booth} The fitting parameters were determined for P=1.4 GPa and held constant for all further pressures; all fitting parameters are listed in Table \ref{table:Fit_param}. After fitting each emission scan, the amplitude of each emission line is integrated as a function of incident energy, and a weighted average is taken to calculate the \textit{f}-occupancy.

Figure \ref{fig:amplitudes} shows typical fits obtained for several incident energies, as well as the obtained amplitudes. Performing this procedure for each pressure point results in a quantitative description of n$_f$ with pressure, as shown in Fig. \ref{fig:nf_dallera}, which agrees with the qualitative description above. The error bars in Fig. \ref{fig:nf_dallera} are derived from the covariance matrix of the least-square fit at the selected emitted energies and line shapes and thus do not include systematic errors; the exact values of n$_f$ shift by approximately $\pm$ 0.1 depending on the line shape and emission energies chosen, but the overall pressure dependence remains consistent. The \textit{f}-occupancy decreases from 1.96 to about 1.76 over the 20 GPa range measured. The transient enhancement near 10 GPa is on the order of 0.15, a significant deviation from the overall trend and a value comparable to the lowest measured pressure. The width of this transient enhancement agrees quite well with structural and transport measurements and they combine to demarcate a sluggish transition region between the competing structures. 

An alternate approach to calculating the valence (or equivalently, n$_f$) can be achieved with a full spectrum fit to the Kramers--Heisenberg equation. We have previously employed this technique for work performed on Yb-compounds under pressure where the energy spacing of the individual absorption peaks was on the order of 10 eV.\cite{Brubaker} In U-compounds, however, the narrow energy spacing of about 4 eV complicates the analysis, and the calculated \textit{f}-occupancy is quite a bit more sensitive to the emission and absorption energies than the method used above. Nonetheless, using the emission energies determined above we could obtain a satisfactory fit to each of the measured spectra, which resulted in slightly lower values of n$_f$, but consistent overall behavior. We have included the results of the Kramers--Heisenberg analysis in the appendix.

\section{Discussion}

Recent manuscripts investigating the \textit{f}-occupation in U-compounds have performed their analysis by holding the energy separation between \textit{f}-states constant at 7.2~eV based on the energy separation between localized \textit{f}$^3$ (UCd$_{11}$) and \textit{f}$^2$ (UF$_4$) materials. \cite{Soderlind, Booth_URS} Using this energy separation for the data presented herein, however, leads to lower quality fits and significant fluctuations in the determined values of n$_f$ for both analysis techniques discussed previously. This could be due to the fact that UF$_4$ was used as a calibrant rather than an intermetallic localized f$^2$ material, which could lead to an artificially large energy spacing, though we point out that we are unaware of any better calibrants. 

We also point out that the same manuscripts no longer include the \textit{f}$^1$ state in their analysis (only \textit{f}$^i$, i=2,3,4,5,6) and the reason is not clear. At least in the case of USb$_2$, it is expected that the f$^1$, f$^2$, and f$^3$ states contribute to the total electronic state in this system. In fact, recent DFT + DMFT calculations that evaluated the competition between local moment physics and electronic itinerancy found that the f$^1$ and f$^3$ states contribute approximately 25\%, which is significantly less than observed in our data, but does support the idea of the multiconfigurational nature of actinide materials. \cite{Miao_NatComm}

The measured \textit{f}-occupation just below n$_f$=2 at the lowest measured pressure is consistent with ambient pressure angle-resolved photoemission spectroscopy data that suggests some of the \textit{f}-electrons have hybridized with the conduction band. \cite{ARPES} As the lattice contracts, this hybridization strengthens, causing the \textit{f}-occupation to decrease as a function of pressure. Interestingly, aside from the transient enhancement, the \textit{f}-occupation decreases semicontinuously across the entire pressure range, seemingly impervious to the different magnetic and crystal structures and the large volume collapse of 17\%. This suggests that the hybridization is insensitive to these factors and may hint at similar local environments of the U-atoms in the low- and high-pressure phases.

The transient enhancement in \textit{f}-occupation across the critical pressure is at odds with the otherwise continuous decrease in \textit{f}-occupation and merits further consideration. In rare-earth materials the individual absorption peaks are well separated and discernible and a change in \textit{f}-occupation can easily be determined by comparing the relative intensities of the absorption peaks. In the case of USb$_2$, both the XAS and RXES data only show a broad peak that shifts to higher incident energies with increasing pressure. Although this can certainly be explained by shifting amplitudes of the individual valence peaks, it is important to consider other factors that could shift the peak position, without influencing the \textit{f}-occupation.  

We are aware of three alternate ways that could shift the peak position despite no change in \textit{f}-occupation. As discussed recently, UO$_2$ and UF$_4$ show different peak positions and shape due to the effects of ligand field splitting, despite both exhibiting the n$_f$=2 configuration. \cite{Tobin} Peak shifts due to differences in covalency have also been observed in complex U$^{5+}$ and U$^{6+}$ organic and hydrate phases and were so severe that the U$^{5+}$ structure actually showed a peak position at lower energy than the U$^{4+}$ UO$_2$ specimen. \cite{U_XAS} Yet other recent measurements on $\alpha$-U demonstrated an increase in absorption energy and peak width as compared to standard, localized f$^2$ and f$^3$ materials due to the delocalized nature of the \textit{f}-electrons. \cite{Soderlind} We do not, however, find that these possibilities offer compelling alternate explanations for the data. 

Significant change in covalency is unlikely to occur across a structural transition, and we discount this as an alternate explanation for the change in white line position. It is true that the degree of localization and crystal field differences between the low- and high-pressure phases would result in a shifted absorption edge, but these would also result in significant changes in peak shape. As shown in the XAS scans (Fig. \ref{fig:XAS}), however, the peak shape remains constant across the transition, and we see no evidence of sudden broadening or narrowing of the absorption edge(s). Nonetheless, to probe for shifting absorption edges, we re-evaluated the valence and fluorescence peak locations at several pressures above and below the magnetostructural transition, while keeping the line shapes constant. We do see a small shift in energy across the transition, but this shift is only on the order of 0.3 eV and does not significantly influence the obtained valence. As such, it is unlikely that the proposed transient \textit{f}-occupation enhancement is an artifact of shifting absorption edges due to changes in (i)~covalency, (ii) degree of localization or (iii) crystal field effects. We also point out that the timescale of RXES measurements is on the order of 1 fs and the \textit{f}-electrons are effectively frozen on these timescales. The measurement thus provides an instantaneous snapshot of the average \textit{f}-occupation that is not influenced by possible slowing down of valence fluctuations. In the absence of any of the aforementioned considerations, we are thus left with the conclusion that the \textit{f}-occupation does, indeed, experience a transient enhancement across the magnetostructural transition. 

We are unaware of other, similar transient valence enhancements reported in the literature. High pressure spectroscopic measurements on other U-compounds have been restricted to XAS measurements, in which it is difficult to extract the \textit{f}-occupation, though the white line position is a valuable comparison for the work presented herein. In the case of UCd$_{11}$, no structural transition is observed, and the white line position increases linearly as a function of pressure. \cite{UCd11} In the cases of UPd$_2$Al$_3$, UC, and UN, structural transitions coincide with changes in the slope of the white line position with respect to pressure, though none of these compounds show the sudden, transient shift observed in our work. \cite{UPd2Al3, UP, UTe} UP, which undergoes structural transitions near P=10 GPa and P=28~GPa, shows a change in white line position across the P=10 GPa transition, but is constant across the latter structural transition. \cite{UP, UP_structure} Similarly, the structural transition observed in UAl$_2$ near 10 GPa does not manifest itself in the observed XAS spectra. \cite{UP} Finally, UTe, which exhibits a coupled magnetostructural (FM--FM) transition in which the low- and high-pressure phases coexist from about 10 GPa to 20 GPa, only shows a change in slope of the white line position with respect to pressure, rather than the exotic transient shift observed in USb$_2$. \cite{UTe, UTe_res, UTe_mag} Evidently, the transient \textit{f}-occupation enhancement in USb$_2$ is unique compared to the aforementioned materials and is a critical component in understanding its magnetostructural transition.

\section{Conclusion}

We have shown that USb$_2$ undergoes a tetragonal-orthorhombic transition near 8 GPa, which coincides with the magnetic AFM--FM transition and results in a surprisingly large volume contraction. Our results suggest that USb$_2$ is mixed-valent, with each of the f$_1$, f$_2$, and f$_3$ states playing a pivotal role in establishing the valence of the USb$_2$ system. The valence just above ambient pressure is close to an effective tetravalent state, though this is increased significantly under pressure. The RXES spectra indicate that the \textit{f}-occupation is particularly sensitive to the coexistence of the low- and high-pressure phases. Outside of this region, the \textit{f}-occupation decreases smoothly with pressure.

\section{Acknowledgements}
This work was performed under LDRD (Tracking Code 18-SI-001) and under the auspices of the US Department of Energy by Lawrence Livermore National Laboratory (LLNL) under Contract No. DE-AC52- 07NA27344. This material is based upon work supported by the National Science Foundation under Grant No. NSF DMR-1609855. Portions of this work were performed at HPCAT (Sector 16), Advanced Photon Source (APS), Argonne National Laboratory. HPCAT operations are supported by DOE-NNSA’s Office of Experimental Sciences. The Advanced Photon Source is a US Department of Energy (DOE) Office of Science User Facility operated for the DOE Office of Science by Argonne National Laboratory under Contract No. DE-AC02-06CH11357.

\section{Appendix}

\begin{figure}[hbtp]
	\centering
	\includegraphics[width=\linewidth]{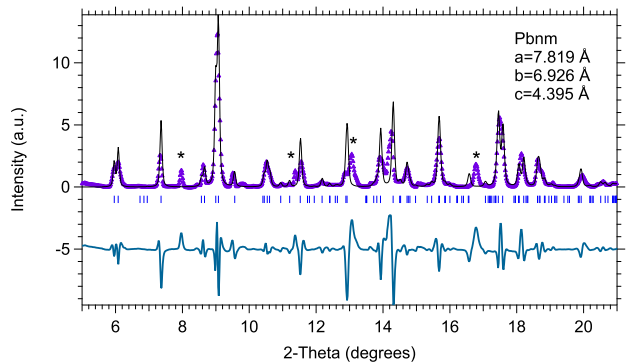}
	\caption{Fit of the high-pressure structure using the Pbnm structure with a=7.819 \AA, b=6.926 \AA, and c=4.395 \AA ~at 10.5~GPa. The reflections are indicated by the blue ticks. Several peaks cannot be indexed using this space group and are indicated with an asterisk (*).}
	\label{fig:PXRD_b}
\end{figure}

Figure \ref{fig:PXRD_b} shows the obtained fit to the high-pressure phase using the Pbnm space group previously determined for UP$_2$ and UAs$_2$. \cite{HTHP} As can be seen, several peaks cannot be accounted for using this structure.

\begin{figure}[hbtp]
	\centering
	\includegraphics[width=\linewidth]{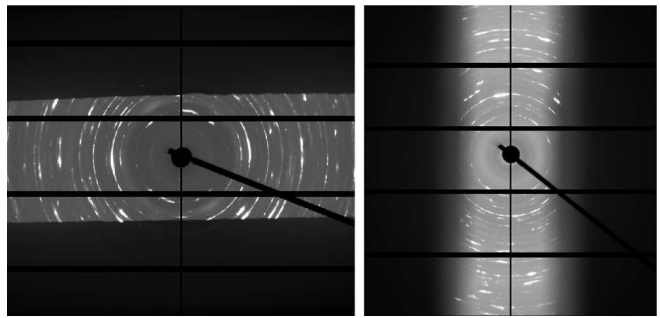}
	\caption{Acquired images for the two distinct measurements of USb$_2$. Left: Image collected at P=10.6 GPa. Right: Image collected at P=10.5 GPa. Both patterns suffer from single-crystal peaks, which make it challenging to extract the atomic positions. The solid black lines correspond to space between individual image plates and the diode, which blocks a section of the emitted X-ray beam.}
	\label{fig:USb2_images}
\end{figure}

Attempts were made to acquire PXRD patterns of the high-pressure USb$_2$ structure with minimal single crystal peaks to confidently extract the atomic positions. For this purpose, a second DAC was prepared with (1) a more carefully ground powder, (2) no rubies, and (3) with a smaller amount of Cu powder, which could be entirely avoided by moving the X-ray beam a small amount within the gasket hole. The data presented in Fig. \ref{fig:PXRD}b are from this DAC. Figure \ref{fig:USb2_images} shows the acquired images for both DACs near P=10.5 GPa. A perfect XRD pattern would consist of constant intensity rings, whereas strong single crystal peaks manifest themselves in select bright spots. As can be seen, both attempts to measure the high-pressure phase of USb$_2$ suffer from strong single crystal peaks. The effect of these single crystal peaks is typically accounted for by using spherical harmonic preferred orientation parameters. However, the peak intensities also depend on the atomic positions, so refining both of these parameters alongside one another can result in significant uncertainties in the atomic positions.

A variety of attempts were made to extract the atomic positions by (1) probing various combinations of Wyckoff positions, (2) using the transformation matrix resulting from the group-subgroup relation between the P4/nmm and Pmm2 space groups to determine a starting point for the atomic positions, and (3) collecting the atomic positions for all materials possessing one of the possible crystal structures with Z=4 and using these as starting points in the Rietveld refinements. These atomic positions were then refined alongside the spherical harmonic preferred orientation parameters to search for a possible solution. In each case, either (i) the obtained fit was unsatisfactory (as judged by residuals) or (ii) the resulting interatomic spacing was unphysical, such as U--Sb or Sb--Sb spacings near 2.25 \AA . In light of these difficulties and unsatisfactory results, we are unable to propose a suitable set of atomic positions. To overcome this obstacle, a theoretical model predicting the high-pressure structure is highly sought-after. We point out that single-crystal diffraction measurements may be able to overcome some of these challenges. However, this measurement can suffer from its own complications, such as single crystals at ambient conditions no longer being single crystals in the high-pressure phase.

\begin{figure}[!hbtp]
\begin{minipage}{\linewidth}
	\centering
	\includegraphics[width=\linewidth]{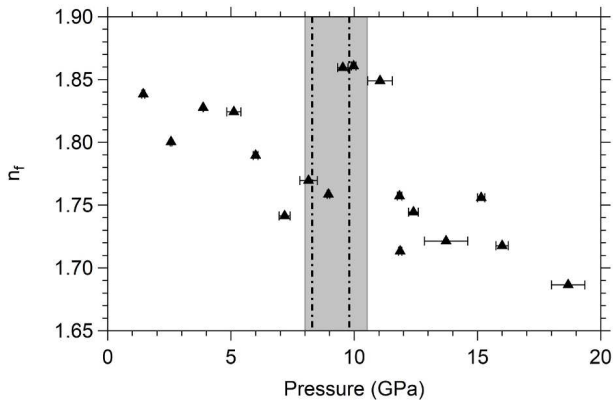}
	\caption{n$_f$ determined via the full spectrum Kramers--Heisenberg analysis. Uncertainty of 1 standard deviation at fixed peak positions is smaller than the markers.}
	\label{fig:KH}
\end{minipage}
\end{figure}

\begin{figure}[hbtp]
	\centering
	\includegraphics[width=0.99\linewidth]{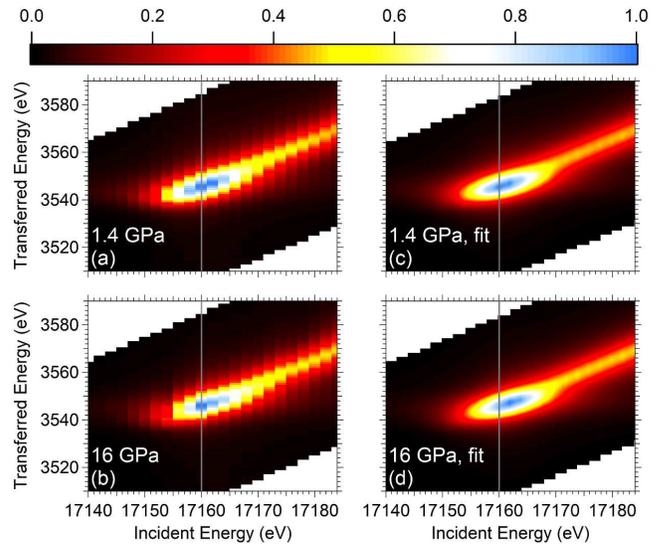}
	\caption{(a--b) RXES spectra and (c--d) fit for P=1.4 GPa and P=16 GPa. The vertical grey line is centered at 17,160~eV and is a guide to the eye.}
	\label{fig:KH_fit}
\end{figure}

Figure \ref{fig:KH} shows the calculated f-occupation using the full-spectrum Kramers--Heisenberg fit. The values are roughly 0.1 lower than those reported in the main text, which is similar to the offset that Booth et al. observed for their RXES measurements on U- and Pu-compounds. \cite{Booth} Nonetheless, the f-occupation shows the same qualitative pressure dependence and a significant enhancement when the low- and high-pressure phases coexist. Table \ref{table:KH} shows the energies that were used for the refinement and Fig. \ref{fig:KH_fit} shows the fit for 1.4 GPa and 16~GPa.

\begin{table}[!ht]
\centering
\renewcommand\arraystretch{1.2}
\caption{Incident and transferred energies used for the full spectrum Kramers--Heisenberg fit.}
\label{table:KH}
\begin{ruledtabular}
\begin{tabular}[t]{lccccc}
E$_{t1}$&E$_{t2}$&E$_{t3}$&E$_{i1}$&E$_{i2}$&E$_{i3}$\\

\hline

3548.4&3544.2&3540.0&17163.1&17159.1&17154.2\\

\end{tabular}
\end{ruledtabular}
\end{table}

\FloatBarrier

\end{document}